\documentclass[epj]{svjour}
\usepackage{epsfig,graphicx,float,pst-all}
\usepackage{booktabs}
\usepackage{amssymb, amsmath}

\newcommand{\be}{\begin{equation}} 
\newcommand{\ee}{\end{equation}}
\newcommand{\bc}{\begin{center}}
\newcommand{\ec}{\end{center}}

\usepackage{booktabs}

\begin{document}
\title{Alpha-induced inelastic scattering and alpha-transfer reactions in $^{12}$C and $^{16}$O within the Algebraic Cluster Model}
\author{Jesus Casal\inst{1,2}\and Lorenzo Fortunato\inst{1,2}\and Edoardo G. Lanza\inst{3,4}\and Andrea Vitturi\inst{1,2} }
%
%
\institute{Dipartimento di Fisica e Astronomia "Galileo Galilei", Universit\`a di Padova, Italy \and INFN Sezione di Padova, Padova, Italy
\and INFN Sezione di Catania, Catania, Italy \and 
Dipartimento di Fisica e di Astronomia "Ettore Majorana", Universit\`a di Catania, Italy}
\date{Received: date / Revised version: date}
%
\abstract{
The molecular algebraic model based on three and four alpha clusters
is used to describe the inelastic scattering of alpha particles populating low-lying states in $^{12}$C and $^{16}$O.  Optical potentials and inelastic formfactors are obtained by folding densities and transition densities obtained within the molecular model.  One-step and multi-step processes can be included in the reaction mechanism calculation.  In spite of the simplicity of the approach the molecular model with rotations and vibrations provides
a reliable description of reactions where $\alpha$-cluster degrees of freedom are involved and
good results are obtained for the excitation of several low-lying states.  Within the same model we briefly discuss the expected selection rules for the $\alpha$-transfer reactions from $^{12}$C and $^{16}$O.  
\PACS{
      {21.60.Gx}{Cluster models}   \and
      {21.60.Fw}{Models based on group theory} \and
      {24.10.-i}{Nuclear reaction models and methods}\and
      {25.55.Ci}{Elastic and inelastic scattering}
     } 
} 
\maketitle
\section{Introduction}
\label{intro}
The clear evidence for alpha-clustering in even and odd light nuclei is one of the most interesting long-standing (but still very hot) features in nuclear structure.  The particular interest arises from the fact that  most of the observables (spectra, transitions, selection rules, one-particle and multi-particle spectroscopic factors, etc) seem to escape conventional descriptions such as the single-particle shell model or the collective
model, showing instead a good agreement with models based on the formation of alpha clusters. A large number of models have been constructed along
the years with several degrees of success that cover various aspects of this complex phenomenology, with the general philosophy of starting from a system of nucleons and looking to which extent the nucleon correlations may lead to clusterization features \cite{kanada}.  A different direction has been taken by the line of research based on the Algebraic Cluster Model (ACM) \cite{Wee,Bij02,Bij14,Bij17,Bij95,Del17,Del17b,Stel,Stel2}, in which one assumes preformed alphas and the structure properties of the system are completely due to the dynamics of these alpha particles (with the possible addition of one or more extra nucleons) within a molecular-like approach.  As notable examples, a nice explanation of most of the
low-energy spectral features of $^{12}$C and $^{16}$O can be obtained assuming  
rotational and vibrational excitations
of an 
equilateral
triangular 
configuration of three alphas in the former case and four alphas at the vertices of a tetrahedron in the latter case.
The expected spectra are schematically shown in the upper frames of Fig.1 and Fig.3 for the two cases.  We refer to ref.\cite{Wee} for the nomenclature of the different
vibrational modes and the corresponding bands.  Note that the sequence of allowed spin and parity in the different bands differs from the usual sequence of rigid rotor, being instead ruled by the $D_{3h}$ and $T_d$ discrete symmetries.

This kind of molecular models have a long history, starting from the seminal paper by Wheeler
in 1937 \cite{Wee}, but has been afterwards neglected, in favor of fully microscopic
approaches. Its strong revival is due to the works by Iachello, Bijker and collaborators in the early 2000's \cite{Bij02,Bij14}.
By using group theory techniques the explicit construction of the corresponding algebras allows the derivation of analytic formulas and selection rules for energy levels and
electromagnetic transition rates.  Similarly one can easily determine matter and charge densities and associated form factors in
electron scattering.  The aim of this paper is to illustrate the use of the ACM for the description of inelastic hadron scattering on $^{12}$C and $^{16}$O.  We will begin by determining densities and transitions densities within the geometrical model, not only for the ground state
band, but also for excited vibrational bands. Then we
will use this information to calculate form factors between different states
and these, in turn, will be used to compute inelastic scattering cross-sections for $\alpha$ on $^{12}$C and $^{16}$O in DWBA. While complicated models including nucleon-nucleon interactions
are certainly more advanced, our main aim is to show
that a simple description in terms of rotations and vibrations of the molecular configurations is already sufficient to yield all
the relevant features of the inelastic process.  We will finally briefly discuss, based on the molecular model, the selection rules for alpha transfer reactions connecting $^{12}$C and $^{16}$O.
 
The selected results presented in this review for the case of $^{12}$C are taken from our ref. \cite{vit19}.  The preliminary results for $^{16}$O, on the other hand, are unpublished and will be part of a forthcoming paper (ref.\cite{cas20}).

\section{ Densities and transition densities within the algebraic molecular model.}
\label{sec:1}
The building blocks of the model are the "preformed" alpha particles. The density of each $\alpha$ particle is taken as a gaussian function:
\be
\rho_\alpha(\vec r)= \Bigr(\frac{\alpha}{\pi}\Bigl)^{3/2} e^{-\alpha r^2}
\ee
with $\alpha=0.56(2)$ fm$^{-2}$ as in Ref. \cite{Del17,Del17b}. 
The three dimensional spherical integral of this function is normalized to 1, therefore one should always multiply by two when dealing with charge-related quantities and multiply by four when dealing with mass-related properties.
\subsection{The case of $^{12}$C}
\label{sec:2}
As already mentioned, the assumed molecular configuration of the ground state of $^{12}$C corresponds to three alpha's at the vertices of an equilateral triangle, each particle displaced of an amount, $\beta$ from the center.  Under this assumption the total density is given by
\be
\rho_{gs}(\vec r, \{ \vec r_k\})= \sum_{k=1}^3 \rho_\alpha(\vec r - \vec r_k)
\label{dez}
\ee
with $\vec r_1=(\beta,\pi/2,0)$, $\vec r_2=(\beta,\pi/2,2\pi/3)$ and  \\
$\vec r_3=(\beta,\pi/2,4\pi/3)$ in spherical polar coordinates $(r,\theta,\phi)$, where the co-latitude is always $\pi/2$ because we have chosen a triangle lying in the $\{xy\}$ plane with the particle labeled as $1$ on the positive $x-$axis.  The constant $\beta$ has been chosen equal to 1.82 fm in order to reproduce  both the ground state radius and the $B(E2)$ to the first excited $2^+$ state. Note that, due to the phenomenological nature of our approach,
the model does not explicitily take into account the effect of the Pauli exclusion principle, that is expected to generate some repulsion between the $\alpha$'s at short distances. This repulsion can be simulated with a change in the effective densities for ovelapping particles, that should not appreciably affect the profile of densities and transition densities on the tails, that is the range important for reaction calculations.

\begin{figure}
\resizebox{0.45\textwidth}{!}{%
\includegraphics{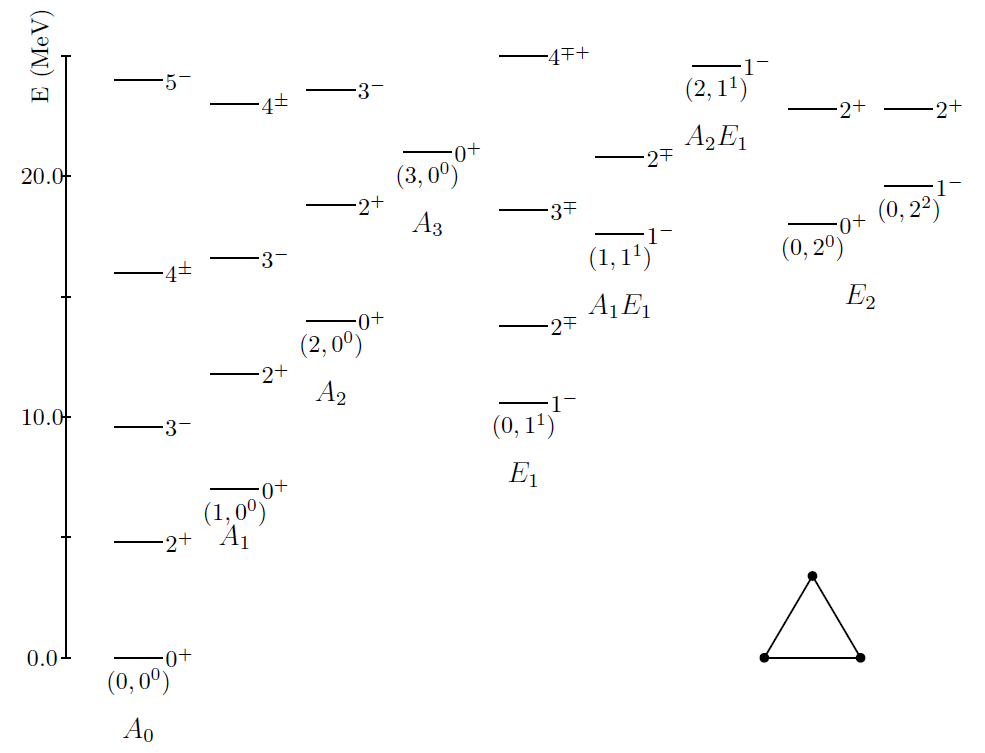}
}
\resizebox{0.55\textwidth}{!}{%
\hspace{0.1cm}
\includegraphics{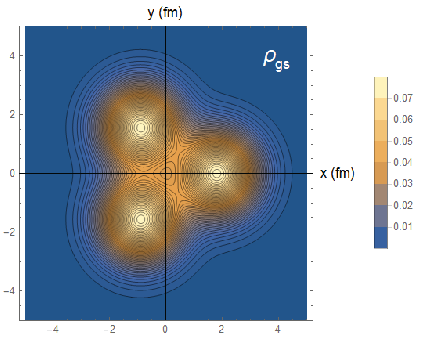}{\centering}
}
\resizebox{0.50\textwidth}{!}{%
\includegraphics{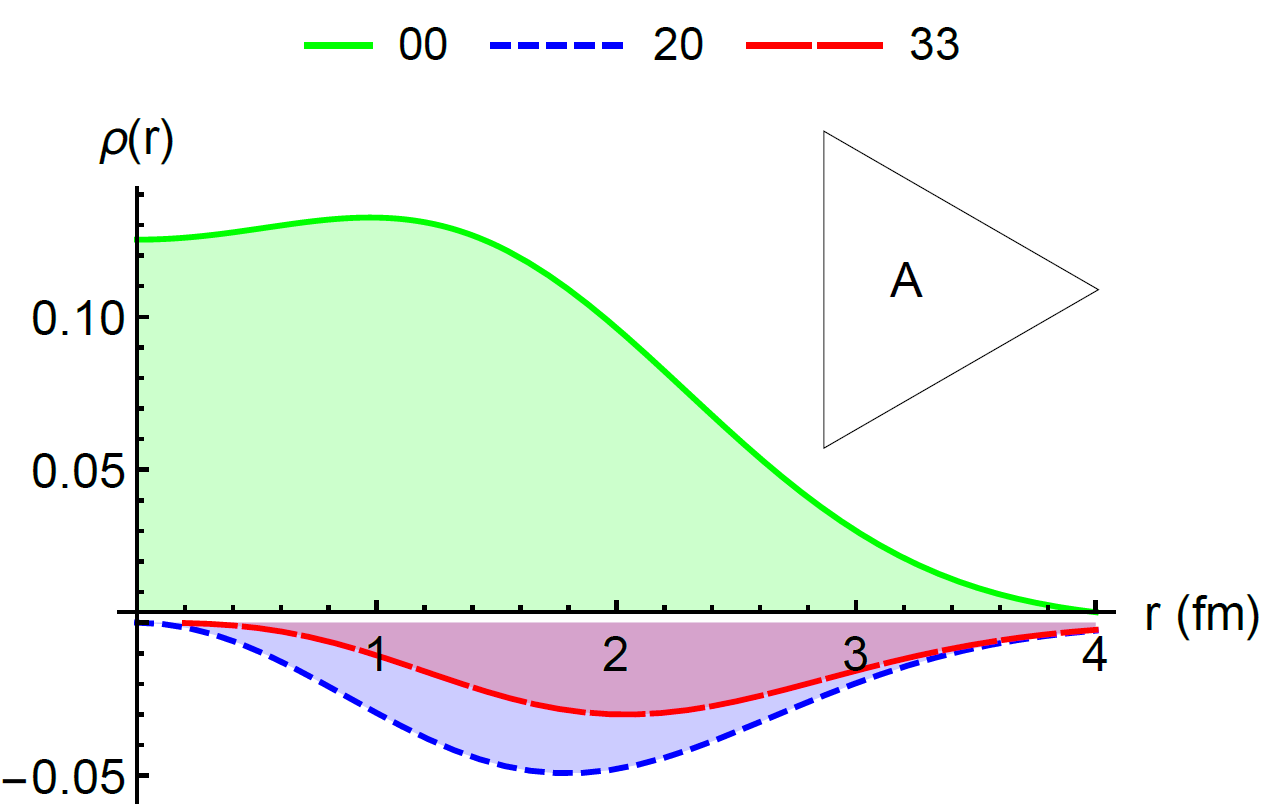}
}
\caption{(Upper frame) Schematic low-lying scheme in $^{12}$C according to the molecular triangular model (adapted from ref. \cite{Bij02}). 
(Central frame) Contour plot of density in fm$^{-3}$ (cut on the $z=0$ plane) of the g.s. static triangular configuration (with A symmetry) in $^{12}$C. 
(Lower frame) Radial transition densities, $\rho_{gs}^{\lambda,\mu}$ within the g.s. band with A symmetry in $^{12}$C}
\label{fig:1}       
\end{figure}

This density is shown as a contour plot in the central frame of Fig.1.   It can be viewed as the density associated to the intrinsic state generating the ground-state rotational band.  Starting from this "intrinsic" density one can determine all densities and transition densities associated to the different members of the rotational band.  This is done by expanding it in spherical harmonics as
\be
\rho_{gs^{12}C}(\vec r)= \sum_{\lambda\mu} \rho_{gs}^{\lambda,\mu} (r) Y_{\lambda,\mu} (\theta,\varphi)
\ee
The different  $\rho_{gs}^{\lambda,\mu}$ are the intrinsic radial transition densities that depend on $\lambda,\mu$ and connect the ground state with the different members of the band. Our choice of coordinates is such that only those multipoles that are allowed by the $D_{3h}$ symmetry appear in the sum.  As examples, we plot in the lower frame of Fig.1 the diagonal density of the 0$^+$ ground state (label 00) and the transition densities to the first two excited members of the band, i.e. the 2$^+$ state (label 20) and the 3$^-$ state (label 33).  Once the densities are known in the intrinsic frame, they can be transformed into the laboratory frame, where the dependence on $\mu$ is lost. Details on how to accomplish this are given for example in Ref.\cite{vit19}. The lab-frame radial transition densities allow the calculation of several intra-band observables, such as the reduced electromagnetic transitions $B(E\lambda)$.  Our aim is to use these densities and transition densities to construct formfactors for the inelastic excitation in alpha-induced reactions  (see next sections).  
\begin{figure}
\resizebox{0.50\textwidth}{!}
{
\hspace{1cm}\includegraphics{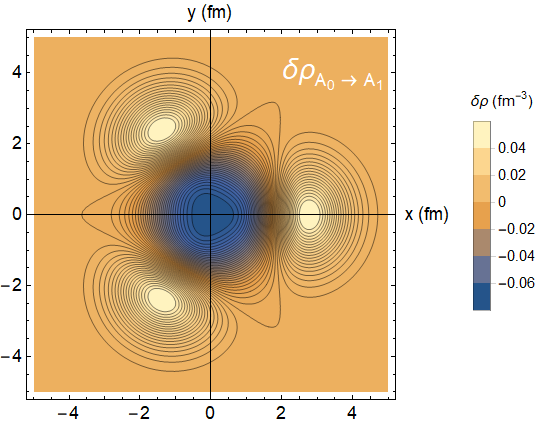}
}
\resizebox{0.45\textwidth}{!}
{\includegraphics{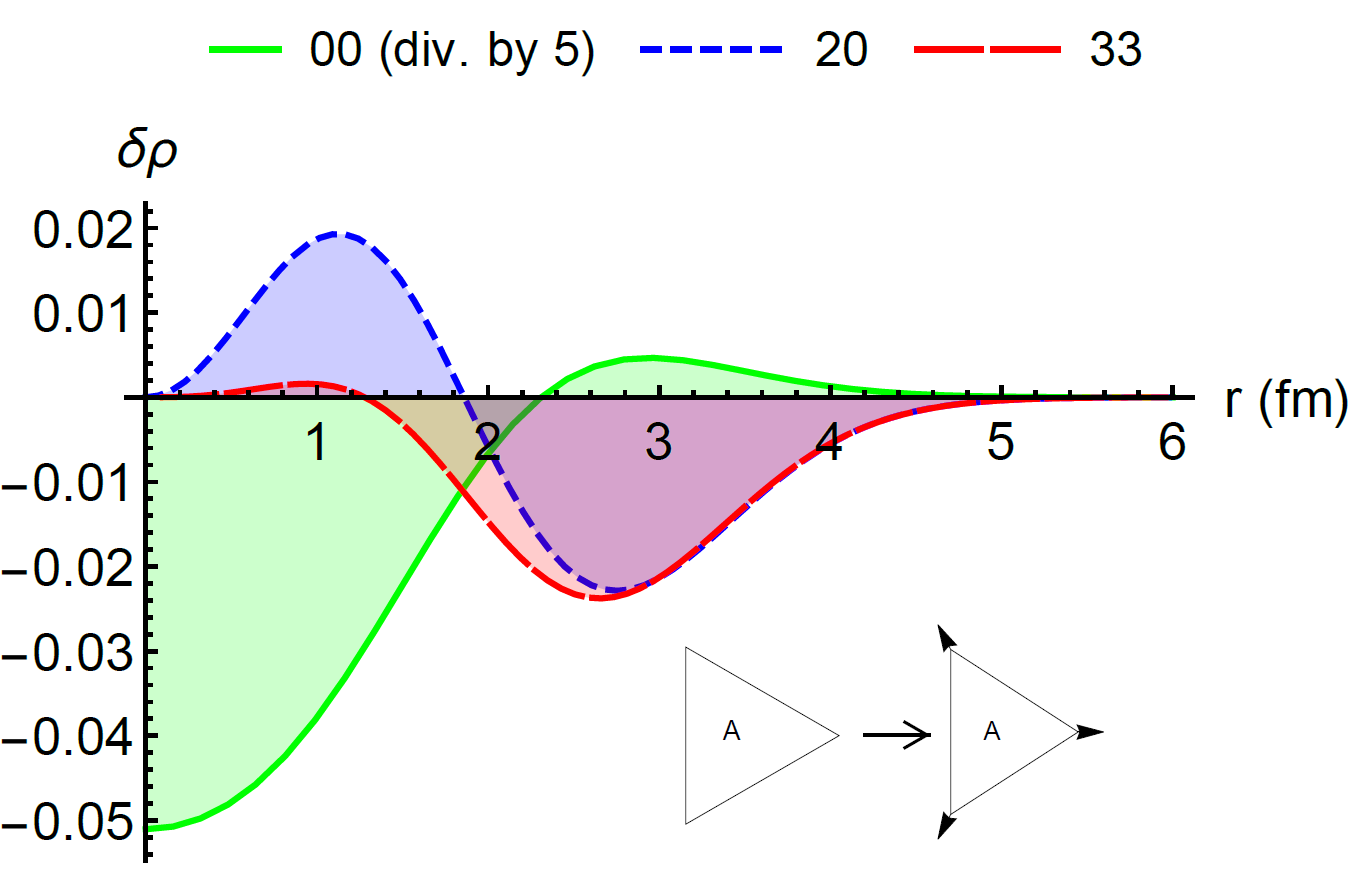}
}
\caption{(upper frame) Transition density from the ground band the first A-type vibration (Hoyle band) in $^{12}$C. (lower frame)
Individual transition densities from the ground 0$^+$ state to the states of the Hoyle band in $^{12}$C, within the expansion in the lowest order spherical harmonics}
\label{fig:2}       
\end{figure}

In analogy with the intrinsic ground state one can then consider the intrinsic states associated with the excited rotational band.  The model assumes that they are associated with vibrations of the three alpha's along the directions of the vectors of normal modes of motion. In the case of $^{12}$C they are are of two types: singly-degenerate fully-symmetric,
A (breathing mode), and doubly-degenerate, E. Each intrinsic state is therefore characterized by the number of different quanta.  The intrinsic state with one A quantum gives rise to the first excited band based on the so-called Hoyle state, of paramount interest for the nucleosyntesis evolution.  The "intrinsic" transition density connecting the ground state band with the Hoyle band can be obtained as an expansion in the small displacements at leading order:
\be
\delta \rho_{gs \rightarrow A} (\vec r) =\chi_1\frac{d}{d\beta}\rho_{gs^{12}C}(\vec r, \beta)~.
\label{chi1}
\ee
In our calculation, we set the intrinsic strength $\chi_1$ to the value  $\chi_1=0.247255$ to reproduce the monopole matrix element $M(E0)$  
connecting the Hoyle and the ground 0$^+$ state, as given in ref. \cite{kan}.   A contour plot (for z=0) of the "intrinsic" transition density is shown in the upper frame of Fig.2. The cut shows the moment at which the
particles oscillate away from the center in a synchronous
fashion, thus depleting the central region (negative tran-
sitions density) and enhancing the external regions (pos-
itive transition density).  This "intrinsic" transition density from the ground-state band to the first excited A-band can be expanded in spherical harmonics in the form:
\be
\delta \rho_{gs \rightarrow A} (\vec r) =\sum_{\lambda\mu}\delta \rho_{gs \rightarrow A}^{\lambda\mu}(r)Y_{\lambda \mu} (\theta,\varphi)
\label{expdrho}
\ee
to yield to individual transition densities from the members of the ground-band and the members of the Hoyle band.  The first allowed and more relevant transition densities are shown in the lower part of Fig.2, corresponding to the transitions from the ground-state 0$^+$ to the Hoyle 0$^+$ state (label 00), to the 2$^+$ state (label 20) and to the 3$^-$ state (label 33) of the Hoyle band, respectively.

With a similar procedure one can determine the other transition densities, for example from the ground band to the E-band, as well as all the in-band transition densities within the different bands.  All details regarding $^{12}$C can be found in ref. \cite{vit19}. 

\subsection{The case of $^{16}$O}
\label{sec:3}
We move now to the case of $^{16}$O system.  In this case the model assumes as ground state configuration four alpha particles at the vertices of a equilateral tetrahedral shape.  As in the previous case of $^{12}$C, excitations of this configuration correspond to vibrations of the alpha according to the normal modes of the systems (labelled A,E and F, according to the nomenclature in ref.\cite{Wee}).  The corresponding rotational vibrational spectrum is schematically shown in the upper frame of Fig.3.  Note that the sequence of allowed states, with angular momentum, parity and degeneracy, differs significantly from the one present in $^{12}$C also for the vibrational bands of type A and E.

As in the previous case the total ground state density can be obtained by summing the density of each alpha particle
\be
\rho_{gs^{16}O}(\vec r, \{ \vec r_k\})= \sum_{k=1}^4 \rho_\alpha(\vec r - \vec r_k)
\label{de16O}
\ee
$\vec r_1=(2\sqrt{2}/3\beta,  0, -\beta/3)$, 
$\vec r_2=(-\sqrt{2}/3\beta, \sqrt{2/3} \beta, -\beta/3)$, \\ $\vec r_3=(-\sqrt{2}/3\beta, -\sqrt{2/3} \beta, -\beta/3)$ and $\vec r_4=(0,0,\beta)$  in cartesian coordinates $(x,y,z)$,
depending on the
parameter $\beta$, the distance of each $\alpha$ particle from the
center of the tetrahedral configuration.  In our calculation this parameter has been assumed equal to 2 fm, according to \cite{Del17}.  The contour plot of the tridimensional density of the g.s. static tetrahedral configuration is displayed in the middle frame of Fig.3.  This "intrinsic" density can then be expanded into multipoles to give the in-band densities and transition densities within the individual states of the ground-band.  The 0$^+$gs density (label 00)
and the allowed transitions to the lowest excited states, 3$^-$ (label 30) and 4$^+$ (label 40), are shown in the lowest frame of Fig.3.  
\begin{figure} 
\resizebox{0.45\textwidth}{!}{%
\includegraphics{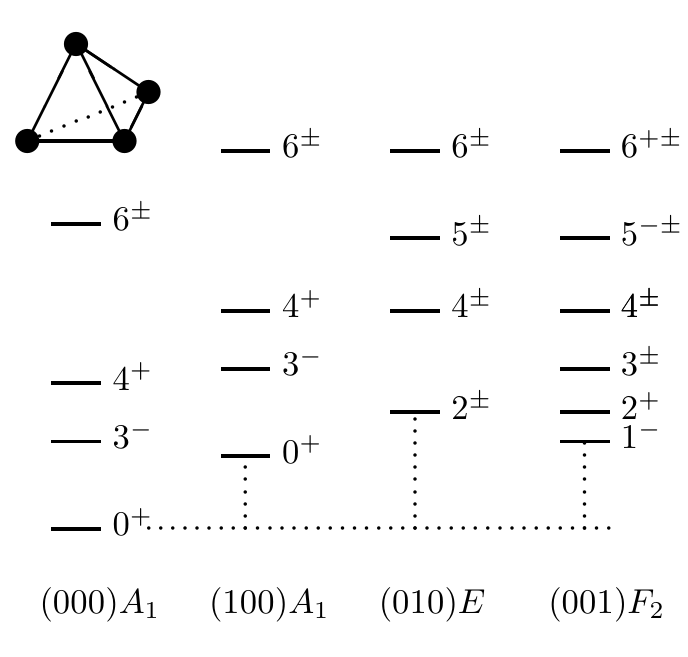}
}
\resizebox{0.50\textwidth}{!}{%
\hspace{1cm}
\includegraphics{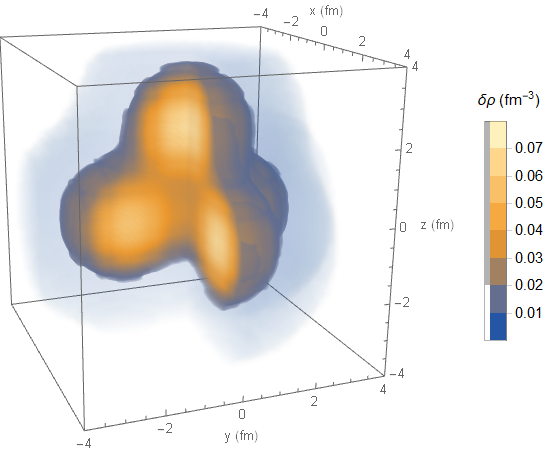}{\centering}
}
\resizebox{0.48\textwidth}{!}{%
\includegraphics{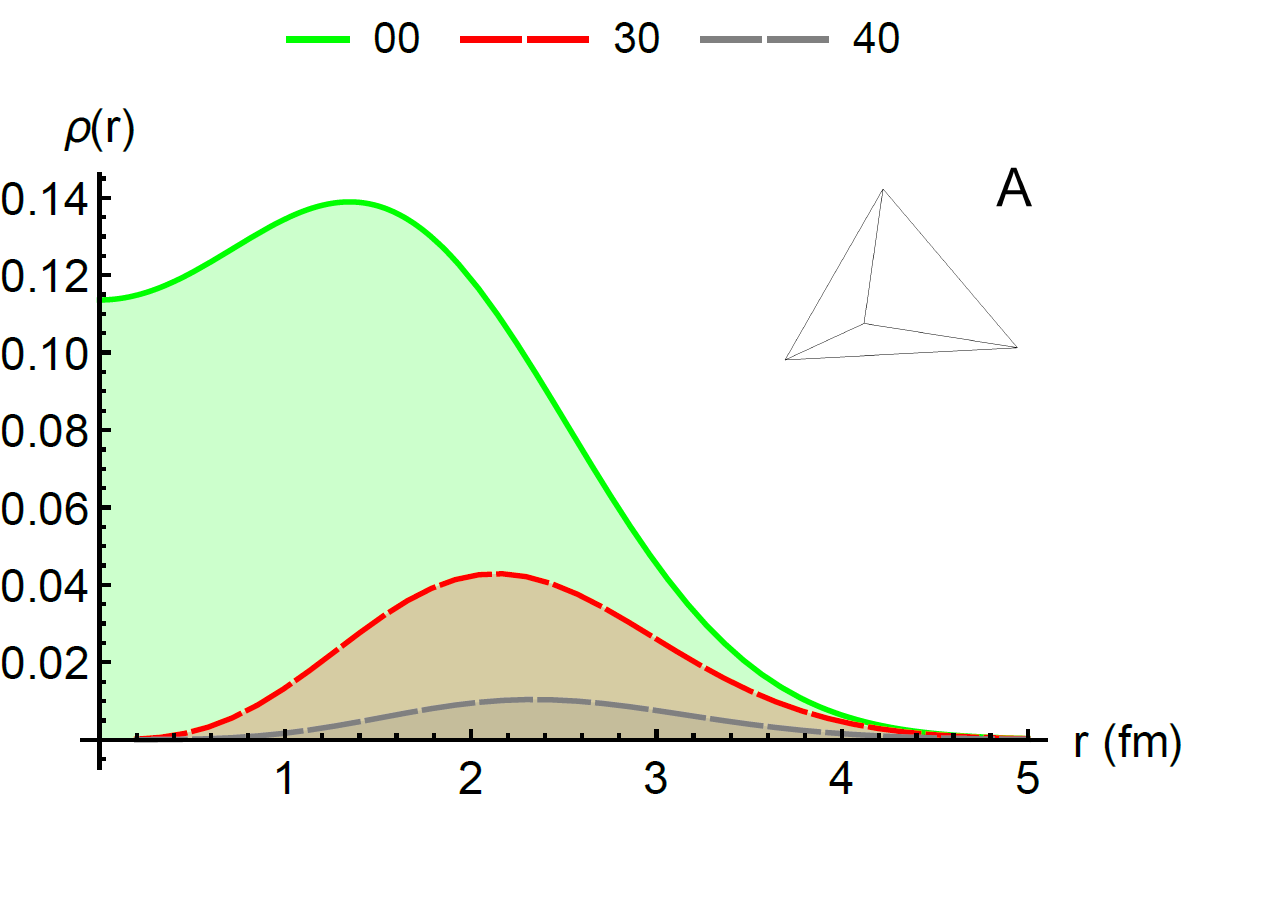}
}
\caption{(Upper frame) Schematic low-lying spectrum of $^{16}$O according to the molecular tetrahedral model (from ref.\cite{Bij17}). 
(Central frame) Contour plot of the 3D density of the g.s. static tetrahedral configuration (with A symmetry) in $^{16}$O. 
(Lower frame) Radial transition densities, $\rho_{gs}^{\lambda,\mu}$ within the g.s. band with A symmetry in $^{16}$O}
\label{fig:3}       
\end{figure}

In analogy to the case of the ground band one can construct the "intrinsic" densities associated with the vibrational modes, and subsequently construct all the corresponding in-band transition densities.  Similarly one can construct the "intrinsic" transition densities connecting, for example, the intrinsic ground state with the intrinsic one-phonon vibrational states.  As an example we can construct in lowest order the "intrinsic" transition density to the first excited symmetric A band in the form
\be
\delta \rho_{gs \rightarrow A} (\vec r) =\chi\frac{d}{d\beta}\rho_{gs^{16}O}(\vec r, \beta)~.
\label{chi1}
\ee
 in terms of the ground state density.  As usual the strenght parameter  $\chi$ has to be chosen, for example, from the experimental value of some intra-band transition matrix element.  In our case the chosen value is 0.22, according to the value of the M(0) matrix element between 0$^+$gs and
the 0$^+$ of the A excited band.  The tridimensional transition density is displayed in the upper figure of Fig.4, showing the typical "breathing-like" behaviour, with opposite signs in the interior and in the external regions (cf. the by-side colour code), separated by a nodal surface.  This is clearly evidenced by the set of intraband transition densities, obtained by multipole expansion of the tridimentional transition density.  In particular the transition density from the ground state 0$^+$ to the 0$^+$ of the A excited band resembles the shape predicted for the breathing modes by collective macroscopic and microscopic models.  

\begin{figure}
\resizebox{0.50\textwidth}{!}
{
\hspace{0.1cm}\includegraphics{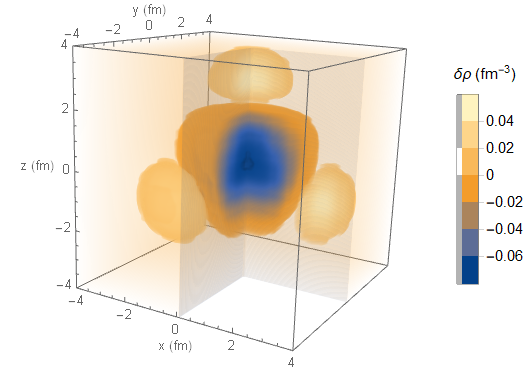}
}
\resizebox{0.48\textwidth}{!}
{\includegraphics{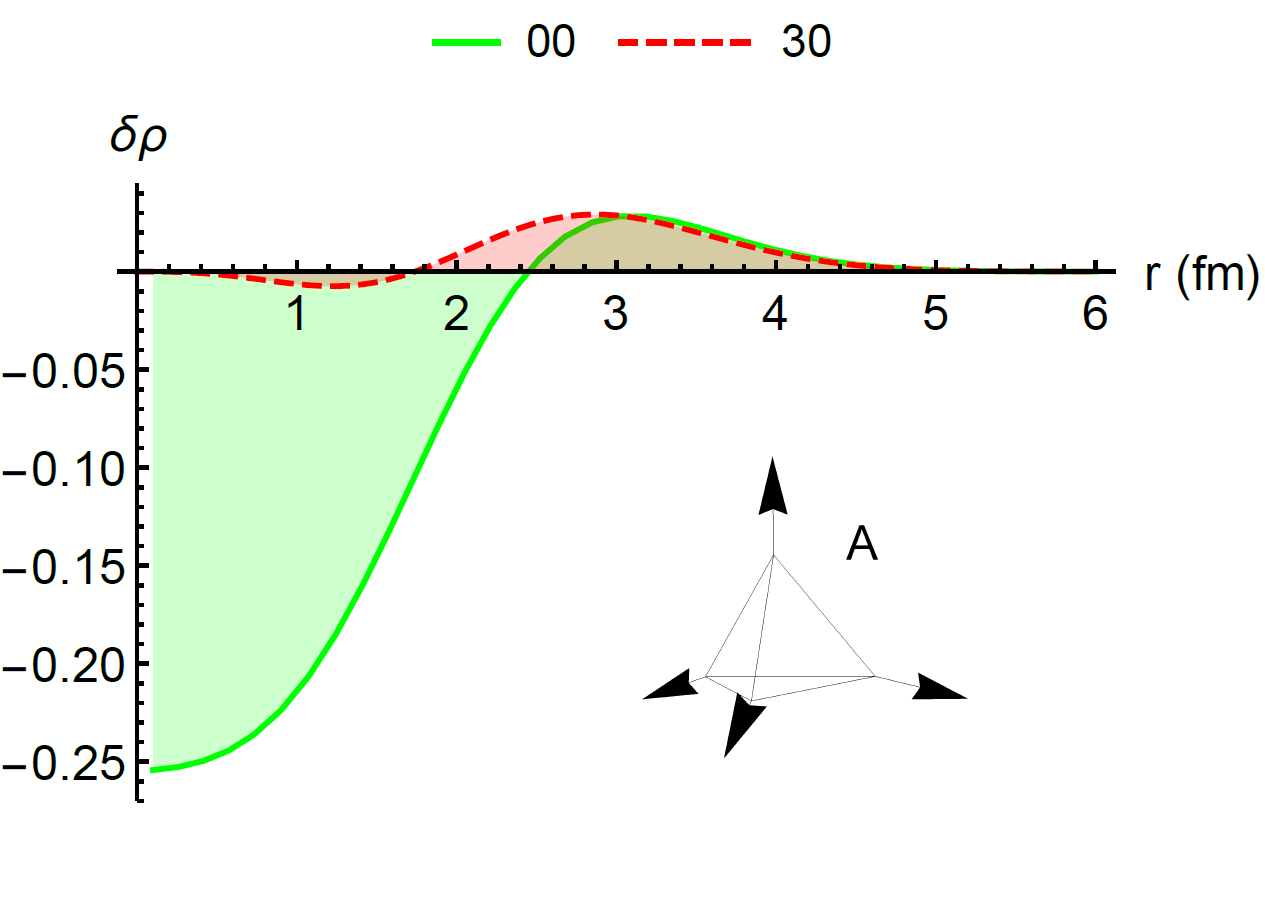}
}
\caption{(upper frame) 3D transition density from the ground band the first A-type vibration in $^{16}$O. (lower frame)
Transition densities from the ground 0$^+$ state to the individual states 0$^+$ and 3$^-$ of the first A-type vibration in $^{16}$O, within the expansion in the lowest-order spherical harmonics.}
\label{fig:4}       
\end{figure}

\section{The formfactors and the description of inelastic scattering induced by alpha particles}
\label{sec:4}
The densities and transitions densities described above in the molecular cluster model contain all the structure information to compute form factors for inelastic excitation processes such as the $\alpha + ~^{12}$C and $\alpha + ~^{16}$O scattering, provided one chooses a suitable nucleon-nucleon potential. We can construct the real part of the nuclear optical model potential using a double-folding prescription as in 
Ref. \cite{sat,sat1}, namely
\be
V_{N}(R)~=~\int\int \rho_{\alpha}(\vec r_1-\vec R) ~\rho_{T}(\vec r_2)~ v_{N}(r_{12})~ d{\vec r_1} d{\vec r_2}
\ee
where $\rho_{\alpha,T}$ are the densities of projectile and target and the effective interaction $v_N$ is a function of the nucleon-nucleon distance $r_{12}$. In this case the $\alpha$ particle is an isoscalar probe ($N=Z$ system), therefore only the isoscalar part of the interaction contributes to the integral. The widely used density dependent Reid type M3Y nucleon-nucleon interaction is used for $v_N$ \cite{sat,m3y}. Examples of optical potentials for the $\alpha + ~^{12}$C case can be found in ref.\cite{vit19}.

Using the transition densities calculated above, one can also compute non-diagonal matrix elements and calculate the form factors by double-folding:
$$F_{ij}(\vec R)= F_{ij}(R) Y_{\lambda \mu}(\hat R) = $$
\be
=\int\int \rho_{\alpha}(\vec r_1-\vec R)~v(r_{12})~ \delta\rho_{i\rightarrow j}(\vec r_2)~ d{\vec r_1} d{\vec r_2}~
\ee
where $v$ contains both the nuclear and coulomb interactions
We show in Fig.5 a compilation of form factors in logarithmic scale, where the nuclear and Coulomb contributions are shown together with the total.
The left sequence refers to the $\alpha + ~^{12}$C case, with the formfactors from the ground 0$^+$ to the 2$^+$ state in the ground band (top panel), to the Hoyle 0$^+$ state (middle panel) and to the 2$^+$ state in the Hoyle band.  The right sequence refers instead to the $\alpha + ~^{16}$O case, with the formfactors from the ground 0$^+$ to the 3$^-$ state in the ground band (top panel), to the 0$^+$ state of the excited A-band (middle panel) and to the 3$^-$ state in the excited band.  Clearly the 0$^+_{gs} \rightarrow $ 0$^+_2$ monopole transitions have only the nuclear part.  The form factors in the two cases are very similar, with the nuclear contribution for the 
$\alpha + ^{16}$O extending to a slightly larger distance due to the larger nuclear radius.
The asymptotic value of the Coulomb contribution depends on the value of the reduced 
transition probability $B(E \lambda)$.

\begin{figure}
\resizebox{0.50\textwidth}{!}
{
\hspace{1cm}\includegraphics{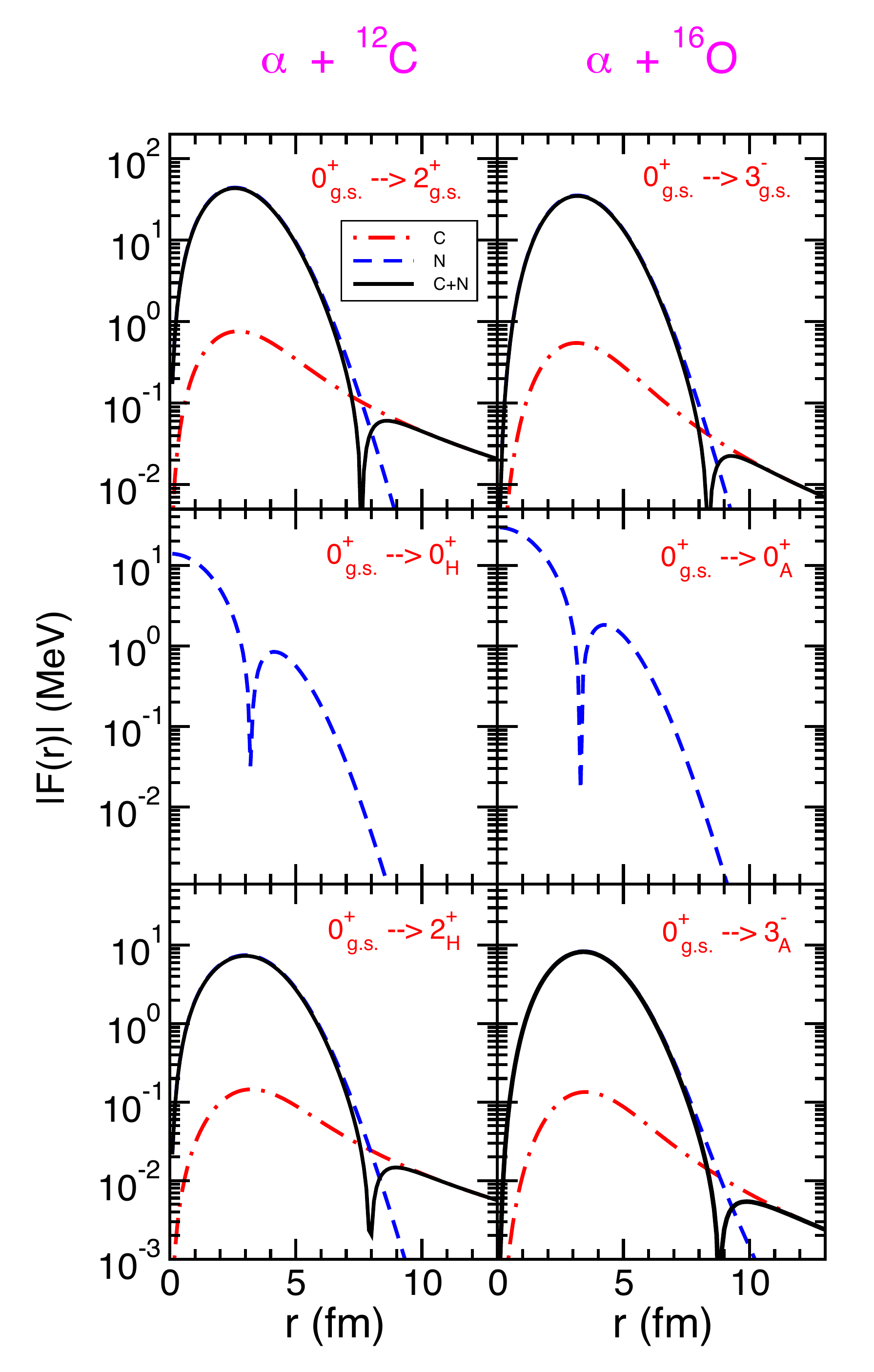}
}
\caption{Form factors in logarithmic scale for a few inelastic
excitation processes of interest. We show the nuclear,
coulomb and total form factors. The left sequence refers to the $\alpha + ~^{12}$C case, while right sequence refers to the $\alpha + ~^{16}$O case. 
}
\label{fig:5}       
\end{figure}


\begin{figure}
\resizebox{0.45\textwidth}{!}
{
\hspace{1cm}\includegraphics{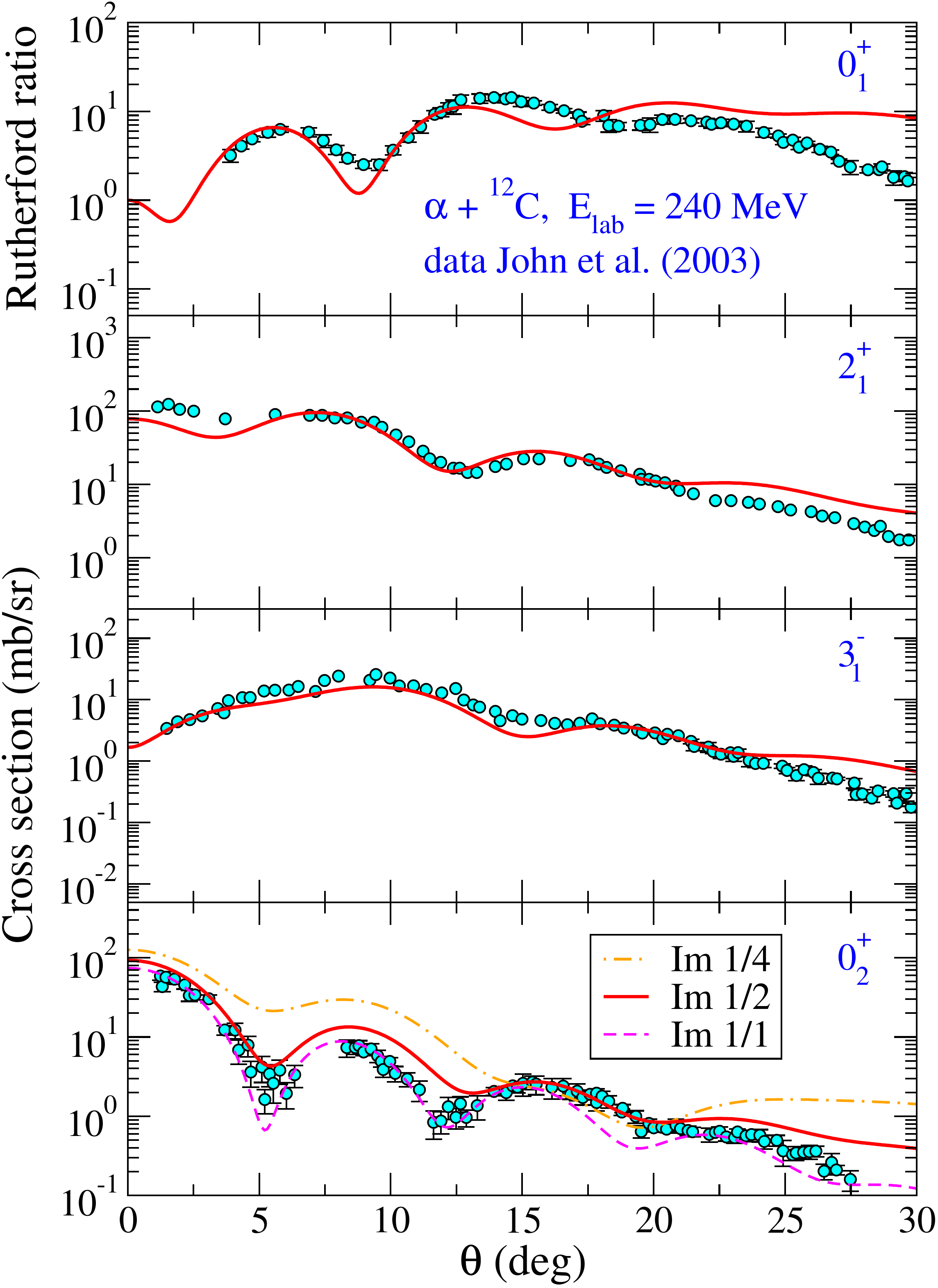}
}
\caption{(upper frames)
Differential cross section for the elastic scattering and
the transitions from the ground state to the 2$^+$ and 3$^-$ states within the ground-state band at 240-MeV bombarding energy for the reaction $\alpha$+$^{12}$C. (lower frame) Differential cross section for the 
population of the Hoyle state at 240-MeV bombarding energy.
  The three curves have different factors for the
depth of the
 imaginary part as indicated in the figure. All data are from Ref. \cite{john}}
\label{fig:6}       
\end{figure}

The potentials and the formfactors evaluated according to the procedure described above can now be used to calculate cross sections for inelastic scattering within a reaction framework (that can be, for example DWBA).  As an example of how the simple geometrical model is
able to capture the main features in reactions where $\alpha$-cluster
degrees of freedom are involved, we present here the computed DWBA
differential cross sections for $\alpha$ + $^{12}$C inelastic scattering. In
the three upper frames of
Fig.6 we show the differential cross section (or ratio to
Rutherford in the elastic case) for the ground-state elastic
scattering and for the excitation to the first excited 2$^+$ and 3$^-$
states within the ground band. In these calculations we
have set the imaginary part of the optical potentials and form
factors to 1/2 of the real part. The bombarding energy is 240 MeV and the corresponding experimental data are from ref.\cite{john}
As apparent from the figure, the $\alpha$-cluster model is able
to reproduce correctly the shape and magnitude of these cross
sections. There is some deviation of the calculated line with
respect to the data at angles above the grazing angle, but
the overall behavior is well reproduced. These results, based on the simple geometrical approach, indicate that alpha clustering plays a vital role and captures the main features of the experimental data. 

In the lowest panel of the figure we give the calculations and data for the transition from the ground to Hoyle 0$^+$ states, exploring the sensitivity to the depth of the imaginary part of the optical potential. The three curves correspond to calculations setting the depth of the imaginary part to different fractions (1/4, 1/2 and 1) of the real part. The theoretical curve is again in rather good agreement with the shape of the experimental data, and the best results are found for values between 1/2 and 1. These calculations indicate that the description of the Hoyle band in terms of breathing vibrations within the geometrical approach is reasonable, finding confirmation in experimental data.

For the population of other states, as for example the 2$^+$ state of the Hoyle band that is expected to be strongly connected to the band head, a description based on the  DWBA  is probably not sufficient, with the direct transition interfering with the two-step process via the Hoyle 0$^+$
(cf. ref. \cite{ito,kan}).  For these aspects we refer to a forthcoming paper \cite{cas20}.

In analogy to the case of $^{12}$C, we have calculated differential cross sections for the 
$\alpha + {^{16}}$O reaction (c.f. [11]). In Fig.7 we show the corresponding DWBA calculations 
for the elastic ($0_1^+$) and the first two inelastic ($0_2^+$,$3_1^-$) channels at 130 MeV. 
As in the previous case of  $^{12}$C, we fixed the imaginary part of the optical potential 
and form factors to 1/2 of the real part. Calculations are compared with the experimental data in 
Ref.~[11]. The agreement for the ground state and the first $3^-$ state is remarkable, particularly 
at small scattering angles, showing that the geometrical model is again able to capture the main 
features of the reaction dynamics. For the elastic channel, our calculations deviate from the data 
as the scattering angle increases, however one should note that no adjustment of the imaginary part 
of the optical potential has been performed. For the $3_1^-$ state belonging to the ground-state band, 
the DWBA result slighly overestimates the data, but the overall shape is well reproduced. In the case
of the population of the $0_2^+$ state, which belongs to the excited A band, the calculations using 
the strength parameter $\chi=0.22$ determined by the known M(E0) matrix element clearly 
overestimate the (scarce) experimental data. If a smaller value (e.g. $\sim$0.11) is employed, the 
results are closer to the data and aligned with the discussion in Ref. [11], where the problem of the 
missing E0 strength is raised. It would be interesting to see whether full coupled-channel calculations, 
including multistep transitions between all low-lying states in $^{16}$O, could improve the agreement and 
explain this discrepancy.

It is also worth noting that, for simplicity, in the case of $0_2^+$ state we used the same optical 
potential in both the entrance and exit channels. As shown in Ref. \cite{vit19}, the 
potential in the excited band should be different due to the possible different mean radius of the 
band head, however fixing this difference would require an additional experimental input (such as a 
precise knowledge of the $0_2^+$ radius or some intraband transition strength). The sensitivity to this 
choice, as well as coupled-channel effects beyond the DWBA approximation, will be subject of further 
investigations. 
\begin{figure}
\resizebox{0.40\textwidth}{!}{%
\includegraphics{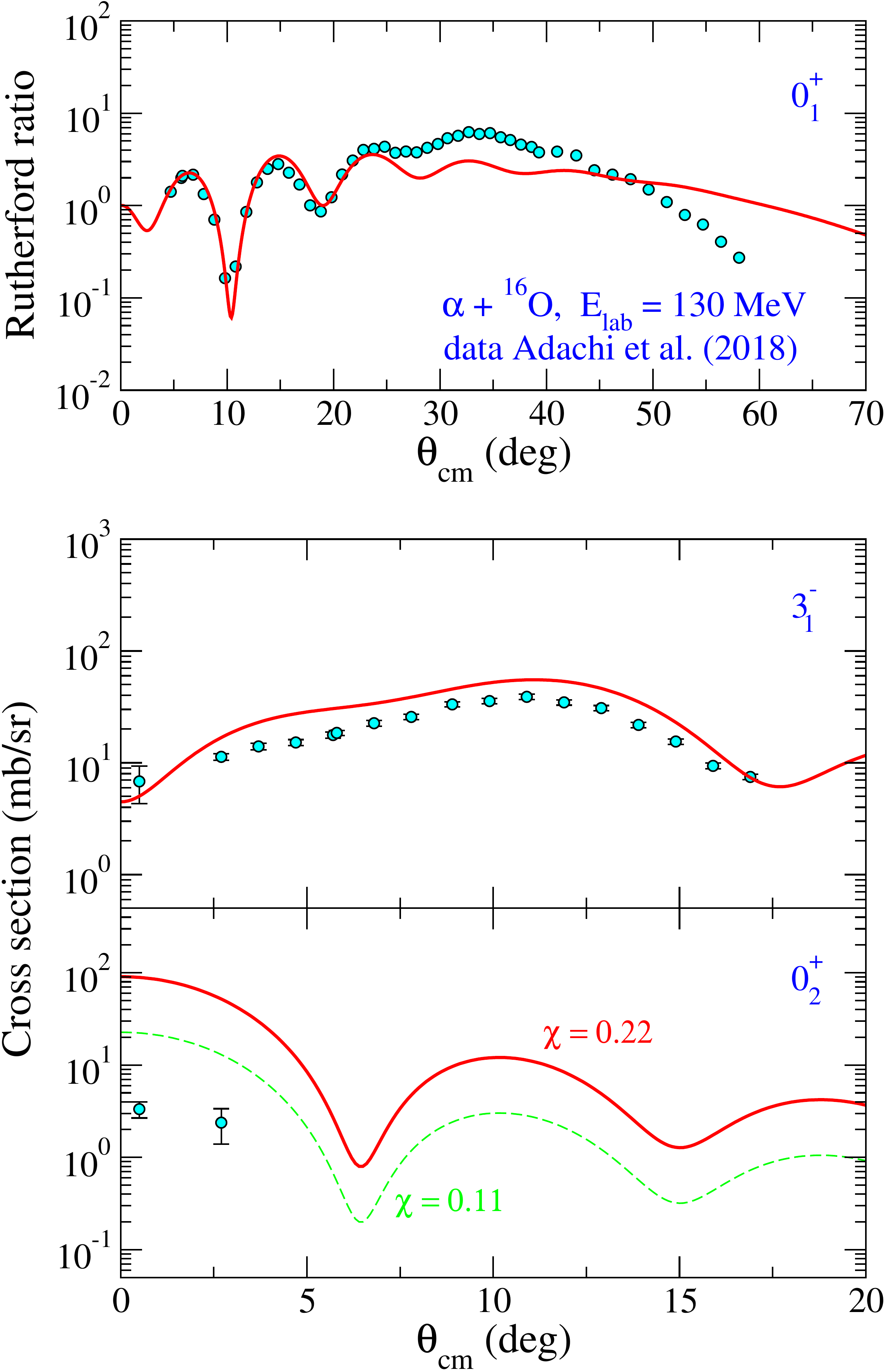}
}
\caption{
Differential cross section for the elastic scattering (top panel) 
and the population of the $3^-$ state within the ground-state band (middle panel) 
and to the $0^+$ state of the excited band (bottom panel) for the reaction $\alpha+{^{16}}$O 
at $E_\alpha=130$ MeV. Experimental data are from Ref. [11]. See the text for details.
}
\label{fig:7}       
\end{figure}
\section{Selection rules for alpha-transfer reactions}

Detailed studies have been performed in the literature concerning the selection rules governing electromagnetic transitions in $^{12}$C \cite{Stel,Stel2} and $^{16}$O, starting from the specific molecular shape associated with the groups ${\cal D}_{3h}$ and ${\cal T}_d$. 
We want to see whether one can derive simple selection rules based on symmetry principles also for the direct $\alpha-$transfer process, i.e. the addition or the removal of an $\alpha$, between two adjacent nuclei in the list of $\alpha-$conjugate ones. The simplest of these reactions is $^8$Be$ +\alpha\rightarrow ^{12}$C and then the series continues with  $^{12}$C$ +\alpha\rightarrow ^{16}$O, etc.

\begin{figure}
\resizebox{0.45\textwidth}{!}
{
\hspace{0.1cm}\includegraphics{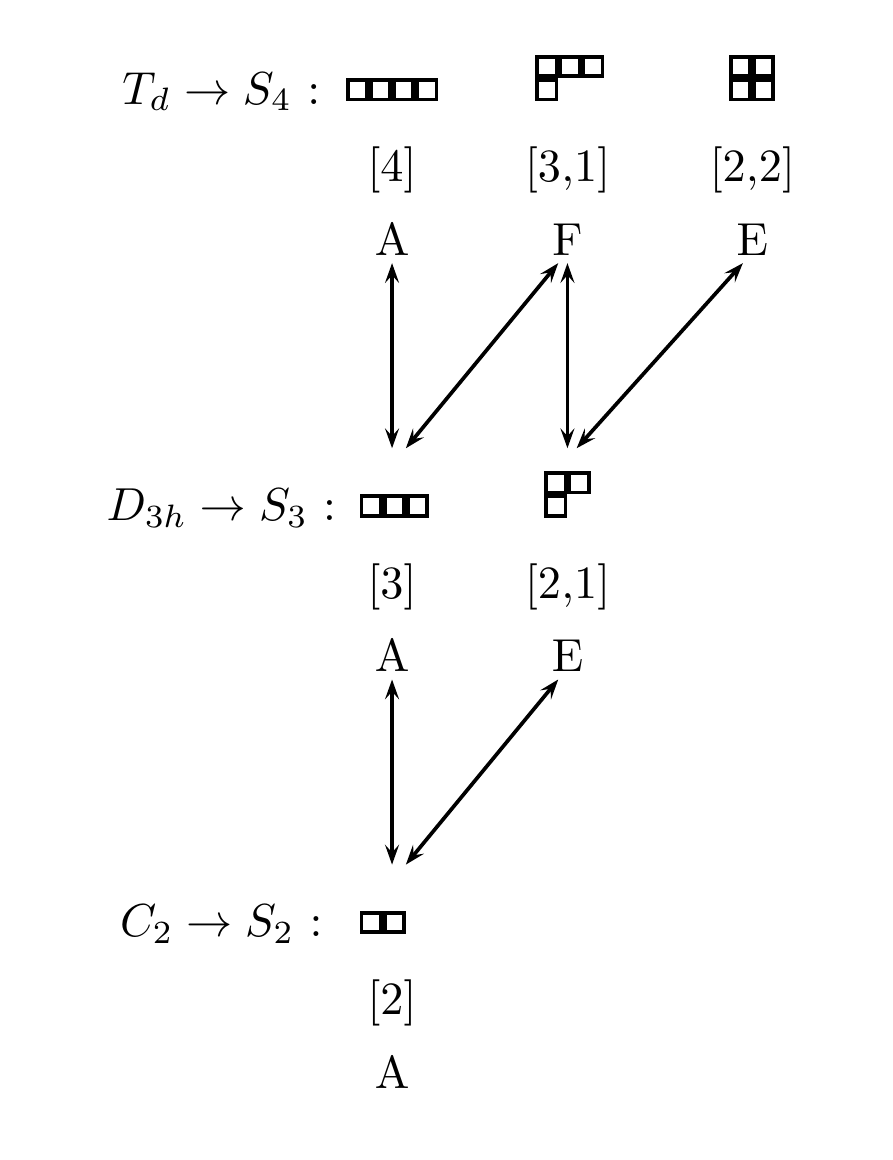}
}
\caption {The representations of the systems with 4,3 and 2 $\alpha$ particles connected by arrows corresponding to processes of induction/restriction that amounts to the addition/removal of one box from the corresponding Young diagrams.
}
\label{sel}       
\end{figure}

These selections rules can be found by connecting the character of the fundamental representations of the underlying discrete group symmetry to those of the corresponding symmetric group S$_n$.  All the $\alpha-$conjugate nuclei are seen as systems of $\alpha$ bosons, therefore the total wave function must be symmetric under the exchange of any two $\alpha$'s.
In addition, if one invokes certain peculiar geometric structures, these systems are also invariant with respect to transformations pertaining to a certain discrete point-group.
As an example in the case of $^8$Be ($N=2$)
the geometric group is ${\cal C}_2$ i.e. the identity and a single $C_2$ rotation.
This group is isomorphic to several other groups 
 \be
{\cal C}_2 \sim C_s\sim C_i\sim Z_2\sim S_2 
\ee 
and in particular to the symmetric group of order 2. The totally symmetric boson states are therefore characterized by the totally symmetric Young tableaux $\Box\Box$. This correlates with the totally symmetric representation A of ${\cal C}_2$.
Similarly for $^{12}$C ($N=3$)
the geometric group is the prismatic group ${\cal D}_{3h}$ that contains the dihedral group $D_3$ that is isomorphic to the symmetric group of order 3, i.e.:
\be
{\cal D}_{3h} \supset {\cal D}_3 \sim C_{3\nu} \sim S_3 
\ee 
The A and E fundamental vibrational modes can be therefore correlated with the totally symmetry Young tableaux and with the mixed-symmetry one.  Finally for $^{16}$O ($N~=~4$) the tetrahedral group is isomorphical to the permutation group of 4 objects
\be
T_{d} \sim S_4
\ee 
and the Young tableaux are correlated to the fundamental A,E and F vibrations as in the first row of Fig. \ref{sel}.

Having established how the various geometric representations correlate with the Young tableaux, we can propose that the direct addition (or subtraction) of an $\alpha$ particle should take the symmetry of the initial and final states into consideration.
Translated in the language of representations of the symmetric group, this means that only Young tableaux that differ for one box (in all possible ways) can be connected by the process of addition/removal of an $\alpha$. This amount to a certain portion, let's call it the "bosonic side" of the full Young lattice.
  
Fig.\ref{sel} shows the representations of the symmetric groups of 4, 3 and 2 particles. Given a certain representation of $S_n$, that can be labeled by the corresponding Young tableau, one can reach induced representations in the larger group or reduced representations in the lower group by adding or removing one box in all possible ways. The arrows indicate the representations that can be connected in this way. Therefore this implies a selection rule for $\alpha-$transfer: two bands, one in $^{12}$C and the other in $^{16}$O, of rotational states built on a certain bandhead with specific vibrational quantum numbers can or cannot be connected depending on the symmetry type to which they belong. For example the ground state band of $^{12}$C and $^{16}$O are both of the A type, therefore the matrix elements of $\alpha$ transfer are allowed by symmetry. Conversely the transfer from the g.s. band or from the Hoyle band of carbon to the E-band of $^{16}$O is forbidden, as it would amount to a reshaping of the Young tableaux that cannot be accomplished with just the addition of one box!

\section{Summary and perspectives}

The molecular cluster model based on "pre-formed" alpha particles has turned out to be able to reproduce spectroscopic properties of the low-energy spectrum of $^{12}$C and $^{16}$O \cite{Wee,Bij02,Bij14,Bij17,Bij95,Del17,Stel}.  The aim of this contribution was to show how the model can be used to describe inelastic excitation of these isotopes, for example in ($\alpha$,$\alpha$') reactions.  In spite of the simplicity of the approach the molecular model with rotations and vibrations provides
a reliable description of reactions where $\alpha$-cluster degrees of freedom are involved and
good results are obtained for the excitation of most states, even within a first-order perturbation treatment.
Further investigations are under process.  In particular we are enquiring the excitation of higher-lying states as well as the role of multistep processes in the reaction mechanism.  We are furthermore investigating
existing data on $\alpha$-transfer reactions from $^{12}$C and $^{16}$O to check the validity of the selection rules that are predicted by the model.    

\section{Acknowledgments}
This work has been partially supported by SID funds (Investimento Strategico di Dipartimento, Università degli Studi di Padova, Italy) under Project No.~CASA\_SID19\_01

\section{Dedication}
This contribution is dedicated to the memory of Mahir Hussein, an exceptional scientist but {\it in primis} a good friend.
%

\end{document}